\documentclass[twocolumn,aps,pra]{revtex4}
\usepackage[]{graphicx}
\usepackage{bm}
\usepackage{amsmath}

\providecommand{\bra}[1]{\langle #1 \rvert}
\providecommand{\ket}[1]{\lvert #1 \rangle}

\begin{document}

\title{Quantum teleportation in the spin-orbit variables of photon pairs}
 
\author{A. Z. Khoury$^1$ and P. Milman$^{2,3}$}

\affiliation{$^1$Instituto de F\'\i sica, Universidade Federal Fluminense,
24210-346 Niter\'oi - RJ, Brasil.}

\affiliation{$^2$Univ. Paris Sud, Institut de Sciences Mol\'eculaires d'Orsay (CNRS),
B\^{a}timent 210--Campus d'Orsay, 91405 Orsay Cedex, France}

\affiliation{$^3$ Laboratoire Mat\'eriaux et Ph\'enom\`enes Quantiques, 
CNRS UMR 7162, Universit\'e Paris Diderot, 75013, Paris, France}

\date{\today}
 
\begin{abstract}

We propose a polarization to orbital angular momentum teleportation scheme using entangled 
photon pairs generated by spontaneous parametric down conversion. By making a joint detection 
of the polarization and angular momentum parity of a single photon, we are able to detect all 
the Bell-states and perform, in principle, perfect teleportation from a discrete to a continuous 
system using minimal resources. The proposed protocol implementation demands experimental 
resources that are currently available in quantum optics laboratories. 
\end{abstract}

\maketitle

Teleportation \cite{TELE} is probably one of the most amazing quantum phenomena relying 
on the existence of entanglement, presenting also direct applications on quantum state 
transmission over long distances.
Briefly, the teleportation protocol can be described as follows : 
Alice (A) and Bob (B) share an entangled state of two qubits, 
so that each one of them receives one qubit, called from now on qubit A and qubit B. In addition to this qubit,
A has another qubit prepared in a given 
(unknown to both parts) state that she wants to teleport to B. This qubit is called qubit S from now on.
To achieve teleportation, A 
makes a joint Bell measurement of qubit S and qubit A. As a result, she finds one of the four possible 
Bell states and transmits this result to B through a classical channel. Depending on the measurement result, 
B applies 
to his qubit one of the three Pauli matrices or simply the identity. After this operation, 
the quantum state of qubit S (which has been destroyed by A's measurement) is reconstructed 
on qubit B.

Different experimental schemes for quantum state teleportation have been reported in the 
literature, using photons \cite{DEMARTINI,telephoton}, trapped ions \cite{teleion} or cavity 
QED systems \cite{teleQED}. Most experimental realizations employ three or more particles or 
subsystems, as it is the case in \cite{telephoton} and \cite{teleQED}. However, by using different 
degrees of freedom of the same particle, one can reduce the number of particles involved \cite{DEMARTINI}. 
The combination of the photon polarization with its spatial degrees of freedom 
has recently led to interesting results such as the demonstration of a topological phase for entangled 
qubits with spin-orbit modes \cite{topol}, proposals of hyperentanglement schemes in parametric 
oscillators \cite{hyper}, and investigation of a spin-orbit Bell inequality \cite{bell1,bell2}. 
Also, spin-orbit photonic devices useful for quantum information protocols have been proposed such 
as cryptography schemes \cite{crypt}, controlled-not (CNOT) gates \cite{cnot} and the so called qplates 
\cite{qplates1,qplates2,qplates3,qplates4}. 
In the present work, we propose a teleportation scheme using two photons produced by spontaneous 
parametric down conversion, which are entangled in orbital angular momentum (OAM). While the proposed 
scheme benefits from the advantages offered by photonic implementations, it also allows for complete 
Bell-state measurement of the spin-orbit degrees of freedom. Another interesting aspect of our proposal 
is that it does not depend on the specific entangled orbital angular momentum state that is shared between 
A and B, relying only on its parity. Different entangled angular momentum states with the same parity 
properties can be used to implement the protocol.  In Ref. \cite{CHEN2009} a teleportation protocol for 
OAM states has been proposed. It presents, nevertheless, several differences from the one we describe 
here, as will be discussed in the following.  

The proposed setup consists of a nonlinear crystal, cut for type I phase match 
so that parametric down-converted photons are produced in the same polarization state. The nonlinear crystal 
is then pumped by a vertically polarized beam, prepared in a Laguerre-Gaussian mode with topological charge $l$. 
Assuming that the phase match condition is satisfied, the down-converted photon pairs are produced with 
horizontal polarization. In terms of OAM conservation, phase match imposes that the sum of 
the down converted charges equals the pump charge \cite{mair,caetano,martinelli,opl}. 
In \cite{steve}, it was shown that the photon pairs are entangled in OAM, and their quantum state 
can be written as
\begin{equation}
\ket{\chi_{0}} = \sum_{m=-\infty}^{+\infty}c_m\,\ket{m,H}_A \ket{l-m,H}_B\;,
\label{chi}
\end{equation}
with $c_{m}=c_{l-m}$. The quantum state given by eq.(\ref{chi}) is clearly entangled in OAM.
Of course, a complete description of the spatial quantum correlations between the twin photons should 
also involve entanglement in the radial indexes of the down-converted Laguerre-Gaussian modes. 
However, the usual measurement setups have finite aperture so that only the lowest radial 
order contributes to the coincidence counts. Therefore, we shall neglect the higher radial orders 
throughout the paper. 

Now, let us suppose that the pump beam is prepared in a Laguerre-Gaussian mode with $l=1$. In this case, 
we can rewrite the entangled state above in components corresponding to even (odd) values of OAM 
for Alice's (Bob's) photon and odd (even) values for Alice's (Bob's) photon in the following way:
\begin{eqnarray}
\ket{\chi_{0}} &=& \sum_{-\infty}^{+\infty} c_{2m}\,\ket{2m,H}_A \ket{1-2m,H}_B
\nonumber\\
&+& \sum_{-\infty}^{+\infty} c_{2m+1}\,\ket{2m+1,H}_A \ket{-2m,H}_B 
\nonumber\\
&=& \sum_{-\infty}^{+\infty} c_{2m}\,\left(|2m,H\rangle_A |1-2m,H\rangle_B \right.
\nonumber\\
&+& \left.\ket{1-2m,H}_A \ket{2m,H}_B\right)\;,
\label{chi2}
\end{eqnarray}
where in the second equality we reordered the summation in the odd-even component by 
making $m\rightarrow -m$ and used $c_{1-q}=c_q$. 
In order to describe the protocol, it will be useful to define the following single 
photon OAM parity states:
\begin{eqnarray}
\ket{E}&=&\sqrt{2}\,\sum_{-\infty}^{+\infty} c_{2m}\,\ket{2m}\;,
\\
\ket{O}&=&\sqrt{2}\,\sum_{-\infty}^{+\infty} c_{2m}\,\ket{1-2m}
=\sqrt{2}\,\sum_{-\infty}^{+\infty} c_{2m+1}\,\ket{2m+1}\;.
\nonumber
\end{eqnarray}

The principle of our proposal is the following: an arbitrary quantum state  is first 
encoded on the polarization of photon A, 
and then teleported to the orbital angular momentum of  photon B  by a complete 
spin-orbit Bell measurement realized on photon A only. The polarization quantum 
state of photon A 
can be prepared  by a sequence of wave plates 
capable to implement a general unitary transformation and produce an arbitrary 
polarization state $|\varphi\rangle\equiv\alpha |H\rangle + \beta |V\rangle$, 
where $\alpha$ and $\beta$ are arbitrary complex coefficients satisfying the 
normalization condition \cite{gadget1,gadget2}. 
After the state preparation scheme, we have a total state of the type: 
\begin{eqnarray}
\ket{\chi} &=& \sum_{-\infty}^{+\infty} c_{2m}\,\left(\ket{2m,\varphi}_A 
\ket{1-2m,H}_B \right.
\nonumber\\
&+& \left.\ket{1-2m,\varphi}_A \ket{2m,H}_B\right)\;.
\label{state}
\end{eqnarray}
It is now useful to define a spin-orbit Bell basis as follows:
\begin{eqnarray}\label{basis}
\ket{\phi^{q}_{\pm}}&=&\frac{1}{\sqrt{2}}\left(\ket{q,H}\pm\ket{1-q,V}\right)
\nonumber\\
\ket{\psi^{q}_{\pm}}&=&\frac{1}{\sqrt{2}}\left(\ket{1-q,H}\pm\ket{q,V}\right)\;.
\label{bell}
\end{eqnarray}

State (\ref{state}), rewritten in the basis (\ref{basis}) gives:
\begin{eqnarray}\label{CHIdenovo}
\ket{\chi}&=&\sum_{m=-\infty}^{+\infty} \frac{c_{2m}}{\sqrt{2}}\, 
\left[\,\ket{\phi^{2m}_{+}}_{A}\,\left(\,\alpha\ket{1-2m}_B+\beta\ket{2m}_B\,\right) 
\right.\nonumber\\
&+&\ket{\phi^{2m}_{-}}_{A}\left(\alpha\ket{1-2m}_B-\beta\ket{2m}_B\right) 
\nonumber\\
&+&\ket{\psi^{2m}_{+}}_{A}\left(\alpha\ket{2m}_B+\beta\ket{1-2m}_B\right) 
\label{chibell}\\
&+&\left.\ket{\psi^{2m}_{-}}_{A}\left(\alpha\ket{2m}_B-\beta\ket{1-2m}_B\right) 
\right]\ket{H}_B
\nonumber
\end{eqnarray}

\begin{figure}
\begin{center} 
\includegraphics[scale=.8]{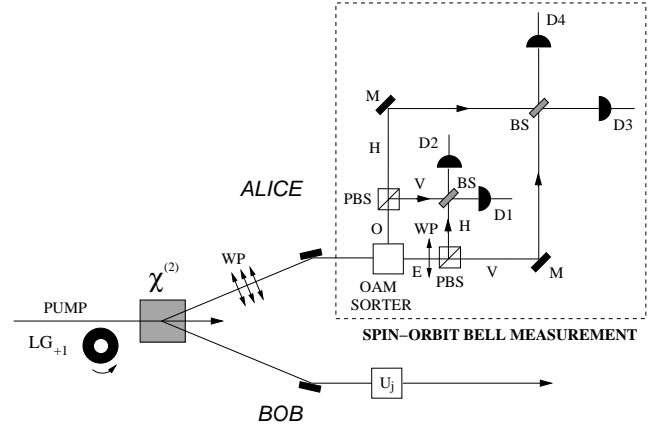}
\end{center} 
\caption{Proposed experimental setup. M - Mirror, WP - Wave Plate, 
BS - Beam Splitter, PBS - Polarizing Beam Splitter, D1-D4 - Detectors.}
\label{fig1}
\end{figure}

Alice can now follow the prescription of \cite{TELE}, as described above, and perform 
a complete Bell measurement on state (\ref{chibell}). Alice's Bell measurement corresponds 
to detecting one of the four maximally entangled state of two different degrees of freedom 
(polarization and OAM) of the same photon. This basis can be completely measured, providing 
a deterministic teleportation protocol, using the set-up sketched in Fig.\ref{fig1}: 
First Alice's photons are sent through an OAM sorter like the one described in 
ref.\cite{sorter}, where the even (E) and odd (O) OAM modes are discriminated in the two outputs. 
Since the $2m$ and $1-2m$ components of the entangled state \ref{chi} have opposite parities, 
they will exit the 
OAM sorter through different outputs. Each output then passes through a polarizing 
beam splitter (PBS) where the $H$ and $V$ components are discriminated. Projection onto 
the spin-orbit Bell basis of Alice's photon is then achieved by recombining the PBS 
outputs in regular beam splitters (BS) and direct photodetection with detectors D1-D4. 
Following the sketch presented in Fig.\ref{fig1}, detectors D1-D4 act as the projectors

\begin{eqnarray}
 P_{\phi_{\pm}}=\sum_{m=-\infty}^{\infty}\ket{\phi^{2m}_{\pm}}\bra{\phi^{2m}_{\pm}}\;, \\ \nonumber
 P_{\psi_{\pm}}=\sum_{m=-\infty}^{\infty}\ket{\psi^{2m}_{\pm}}\bra{\psi^{2m}_{\pm}}\;.
\end{eqnarray}

Alice's apparatus measures 
only the parity of the OAM state, so that an infinity of different entangled OAM states with the same 
parity properties equally fit for the teleportation protocol to work. Photon B, after the measurement, 
is left in 
a superposition of all possible values of the OAM that respect the parity conditions. The relative weights 
of the odd and even components in Bob's photon state are given by the coefficients 
$\alpha$ and $\beta$ of the state Alice wants to teleport. 

In order to complete the teleportation protocol, we allow for the exchange of two classical bits of 
information between A and B. 
A then tells B which one of the Bell states she detected. 
Depending on which of Alice's detectors clicked, Bob's photon is left in one 
of the following states:
\begin{eqnarray}
(\,\alpha\ket{E}&+&\beta\ket{O}\,)\,\ket{H}\;,
\nonumber\\
(\,\alpha\ket{E}&-&\beta\ket{O}\,)\,\ket{H}\;,
\nonumber\\
(\,\alpha\ket{O}&+&\beta\ket{E}\,)\,\ket{H}\;,
\label{bobstate}\\
(\,\alpha\ket{O}&-&\beta\ket{E}\,)\,\ket{H}\;,
\nonumber
\end{eqnarray}
Using the information A provides, B applies an unitary transformation 
$U_j$ (one of the three Pauli matrices, or the identity) to photon B  to 
reconstruct the state $\ket{\varphi}$, photon's A initial state, in photon B, and resume teleportation. 
In this case, 
Alice's polarization state is teleported to Bob's OAM parity state. 
The required unitary transformations in the $\{\ket{E},\ket{O}\}$ subspace can be achieved 
by means of simple optical setups. A Dove prism (DP) performs an image reflection making 
$\ket{m}\rightarrow\ket{-m}$. 
A spiral phase hologram (SPH) adds one unit of OAM to an incoming beam, so that 
$\ket{m}\rightarrow\ket{m+1}$ \cite{MASK1,MASK2,moire}. Since $c_m=c_{1-m}$ one can easily see that 
a DP followed by an SPH makes the transformations $\ket{E}\rightarrow\ket{O}$ 
and $\ket{O}\rightarrow\ket{E}$. Also, any relative phase can be introduced between 
the even and odd components with an OAM sorter followed by a delay line. These 
resources allow Bob to implement the unitary transformation needed to resume 
the protocol. 

\begin{figure}
\begin{center} 
\includegraphics[scale=.45]{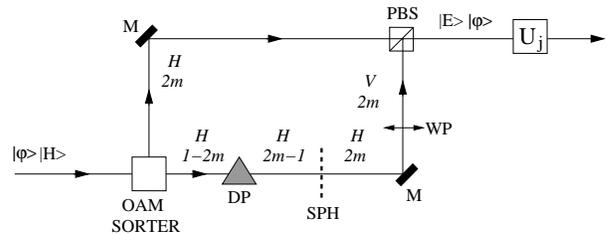}
\end{center} 
\caption{Scheme for OAM parity - polarization swap. M - Mirror, DP - Dove Prism, 
SPH - Spiral Phase Hologram, WP - Wave Plate, PBS - Polarizing Beam Splitter.}
\label{fig2}
\end{figure}

One interesting remark about the protocol is that the only condition imposed to the OAM entangled 
states is that $c_m=c_{1-m}$. Therefore, the whole protocol does not depend on details of the 
state created by parametric down conversion, relying only on its symmetry 
properties. This is why the type I phase match is more adapted to this protocol. 
The action of birefringence on photon pairs created under type II phase match 
would spoil this symmetry and affect the teleported state. Starting from any one of states (\ref{bobstate}), 
Bob can 
reconstruct $\ket{\varphi}$ 
with the help of the aforementioned simple optical devices. As a result, one 
teleports a discrete polarization state to the OAM parity of the single photon 
wavefront. Parity is an usual dichotomization of continuous variables \cite{banaszeck}. 
In our protocol it allows for the quantum state teleportation from a discrete degree 
of freedom to a continuous one. 
Another possibility is to 
swap the OAM parity state to the polarization (see Fig.\ref{fig2}) and then make the 
necessary unitary transformation with polarization devices only. This procedure would 
simplify the tomography of the teleported quantum state. 

It is important to notice the crucial role played by the OAM pump in the 
teleportation scheme. For any value of $l\neq 0$, the dominant term in the 
expansion in Eq.(\ref{chi}) is the maximally entangled component 
$\ket{l}_A\ket{0}_B+\ket{0}_A\ket{l}_B$. Moreover, all the rest of the 
expansion can be cast in the form of a superposition of maximally entangled (ME)
components as in Eq.(\ref{chi2}). Our teleportation scheme works simultaneously 
in each ME states subspace. However, in the absence of the OAM pump ($l=0$), the dominant 
term in (\ref{chi}) will be the product component $\ket{0}_A\ket{0}_B$, and the teleportation 
protocol would not work. 

We briefly discuss now the main differences between the present teleportation scheme 
and the one in \cite{CHEN2009}, where the same goal is pursued: teleporting an OAM state. 
A first difference is that, while in \cite{CHEN2009} the dichotomization of the OAM Hilbert 
state is done using the state's helicity, we use here it's parity. Consequently, the 
experimental apparatus is completely different: while here we entangle parity and 
polarization with the help of beam splitter, in \cite{CHEN2009} entanglement between 
different helicities and polarization is achieved by pumping crystals serving as 
spiral phase plates. A second difference is that in \cite{CHEN2009}, the nonlinear 
crystal responsible for the photon pair generation is pumped by a beam with zero 
angular momentum. As a consequence, the highest probability is to create photon pairs 
with both null OAM. In the present proposal, we increase the proportion of parity 
entangled photon pairs by pumping the non linear crystal with a beam with $m=1$. 
Finally, a crucial difference between both schemes is that the Bell measurement we 
propose is performed in two degrees of freedom of a same photon, so that a total 
number of two photons only is used. In \cite{CHEN2009}, three photons are used and 
the Bell measurement is performed on the polarization of two of them.  

As a conclusion, we have proposed a scheme to teleport the quantum state of 
a two-dimensional variable (polarization) to another one belonging to an infinite 
dimensional space (OAM) using two photons only. Our proposal is possible 
by dichotomizing the infinite dimensional OAM state space and making a Bell 
measurement on the two degrees of freedom of the same photon. It demands 
experimental resources already available in laboratories and can be realized in 
a short delay.

\begin{acknowledgments}
Funding was provided by the Coordena\c c\~{a}o de Aperfei\c coamento de 
Pessoal de N\'\i vel Superior (CAPES), Comit\'e Fran\c cais d'Evaluation de 
la Cooperation Universitaire avec le Br\'esil (COFECUB), 
Funda\c c\~{a}o de Amparo \`{a} Pesquisa do Estado do Rio de Janeiro (FAPERJ), 
and Instituto Nacional de Ci\^encia e Tecnologia de Informa\c{c}\~ao Qu\^antica 
(INCT/IQ-CNPq). 
\end{acknowledgments}

\end{document}